\documentclass[a4paper,onecolumn,english,british]{article}
\usepackage[T1]{fontenc}
\usepackage[latin9]{inputenc}
\usepackage{array}
\usepackage{wrapfig}
\usepackage{url}
\usepackage{amstext}
\usepackage{graphicx}

\makeatletter

\providecommand{\tabularnewline}{\\}

\usepackage{ae}
\usepackage{aecompl}

\usepackage{tikz}

\newcommand{\UG}{{University of Glasgow, UK}}
\newcommand{\email}[1]{{#1}}

\author{ 
  Benjamin Piwowarski  \\ 
 \email{benjamin@bpiwowar.net} 
 \and
 Ingo Frommholz \\ 
 \email{ingo@dcs.gla.ac.uk}
 \and 
 \and Mounia Lalmas \\ 
 \UG \\
 \email{mounia@acm.org}
\and
 Keith van Rijsbergen \\ 
 \UG \\
 \email{keith@dcs.gla.ac.uk}
}

\@ifundefined{showcaptionsetup}{}{%
 \PassOptionsToPackage{caption=false}{subfig}}
\usepackage{subfig}
\makeatother

\usepackage{babel}

\date{}

\begin{document}

\title{Exploring a Multidimensional Representation of Documents and
  Queries~(extended version)%
\footnote{This an extended version of a paper published in RIAO 2010~\cite{Piwowarski2010Exploring-a-Multidimensional}.%
}}
\maketitle
\begin{abstract}
In Information Retrieval~(IR), whether implicitly or explicitly,
queries and documents are often represented as vectors. However, it
may be more beneficial to consider documents and/or queries as multidimensional
objects. Our belief is this would allow building {}``truly'' interactive
IR systems, i.e., where interaction is fully incorporated in the IR
framework.

The probabilistic formalism of quantum physics re\-pre\-sents events
and densities as multidimensional objects. This paper presents our
first step towards building an interactive IR framework upon this
formalism, by stating how the first interaction of the retrieval process,
when the user types a query, can be formalised. Our framework depends on a number of parameters affecting the final
document ranking. In this paper we experimentally investigate the
effect of these parameters, showing that the proposed representation
of documents and queries as multidimensional objects can compete with
standard approaches, with the additional prospect to be applied to
interactive retrieval.

\end{abstract}
\global\long\def\bra#1{#1^{\top}}
\global\long\def\ket#1{#1}
\global\long\def\braAket#1#2#3{#1^{\top}#2#3}

\global\long\def\braket#1#2{#1^{\top}#2}
\global\long\def\pr{\mbox{Pr}}
\global\long\def\event#1{\mathbf{#1}}
\global\long\def\kbasis#1{\ket{#1}}
\global\long\def\bbasis#1{\bra{#1}}

\global\long\def\norm#1{\left\Vert #1\right\Vert }
 \global\long\def\vproj#1{\ket{#1}\bra{#1}}

\global\long\def\tr{\mathrm{\mathrm{tr}}}
\global\long\def\card#1{\mbox{card}\left(#1\right)}

\global\long\def\units{\mathcal{U}}

\section{Introduction}

Most information retrieval (IR) models, including probabilistic and
vector ones, use the same underlying one-dimensional representation
of documents and que\-ries, i.e., as vectors defined in a vector
space, typically a term space. However, this representation has some
limits when dealing with more complex IR aspects like interaction,
diversity and novelty%
\footnote{In our research, we are particularly interested in these aspects of
the IR process.%
}. Indeed, recent research showed that these complex aspects of the
retrieval process benefit from more sophisticated representations
of documents and queries~\cite{Wang2008A-study-of-methods,Che2006A-stereo-document},
in particular those providing for more powerful geometric manipulations
of IR components.

The representation of documents and queries in IR should evolve so
the user interaction can be incorporated in a natural and principled
way in the IR process~\cite{Rijsbergen2004The-Geometry-of-Information}.
Our claim is that representing documents and queries as \emph{multidimensional}
objects~(e.g. subspaces in a vector space) allows for not only a
novel but also a more powerful way to tackle this challenge. This
representation is particularly interesting from a theoretical point
of view because it is possible to use a principled interpretation
of the probabilities associated with such multidimensional objects,
which comes from quantum physics~\cite{Rijsbergen2004The-Geometry-of-Information}~--
the so-called {}``quantum probabilities'' framework. This representation
is also interesting from an intuitive point of view because it relies
on a geometric representation of documents and queries in a vector
space, which has proved successful in IR~\cite{Baeza-Yates1999Modern-Information}.
This representation reveals also a strong connection between orthogonality~(in
the vector space) and non-relevance, which has been successfully used
to represent term negation in queries~\cite{Widdows2003Orthogonal-negation}.

In~\cite{Piwowarski2009A-Quantum-based-Model}, a framework for interactive
IR that relies on such a multidimensional representation of documents
and queries was proposed. In this framework, the user's information
need~(IN) is represented by a set of weighted vectors that evolve
with the user's interaction. A probability of relevance of a document
(for that IN) is computed with respect to this set. Although the components
of our framework were described, they remained abstract. In particular,
no explicit document and query representations were proposed. The
next step is to operationalise the framework, which is the focus of
this paper. We show how document and query representations are computed
to then allow estimating the probability of relevance of the document
to a given IN.

With respect to related work, multidimensional representations, respectively,
of queries were used in~\cite{Wang2008A-study-of-methods} to model
negative user feedback, and of documents were investigated in~\cite{Che2006A-stereo-document}
in an ad hoc setting. Our work encompasses those since it provides
a principled and probabilistic way to work with multidimensional objects.
Finally, two lines of research explored, respectively, a subspace
representation of documents~\cite{Melucci2008A-basis-for-information}
and of a user's IN~\cite{Melucci2008A-basis-for-information}. In
our work, we go further and show that both documents and INs can be
represented as multidimensional objects, and propose a principled
methodology to construct these representations.

The outline of this paper is as follows. We first briefly introduce
our framework and describe how the probability of relevance is computed
within the quantum probability framework (Section~\ref{sec:framework}).
Next we show how we construct the query and document representations,
and introduce several parameters for these representations (Sections~\ref{sub:document-representation}
and~\ref{sub:Creating-the-Query}). Finally, we present experimental results, which validate our document
and query representations, some of the investigated parameters, and
give insights on how our framework can be further developed (Section~\ref{sec:experiments}).

\section{A Quantum-inspired View for IR}

\label{sec:framework}

Our IR framework is built upon~\cite{Piwowarski2009A-Quantum-based-Model},
which is based on quantum probabilities and where we assume that there
exists a vector space of \emph{pure}%
\footnote{The concept of {}``pure'' IN is new and central to our framework.
In this paper, we use {}``pure IN'' to distinguish it from {}``IN'',
where the latter refers to information need in its usual sense in
IR, e.g., see \cite{Ingwersen2005The-Turn-Integration}.%
}\emph{ information needs}~(INs), where each vector corresponds to
an IN that completely characterises a possible user's IN~--~by analogy
with quantum physics where a vector completly characterises a physical
system. Knowing a user's pure IN would determine which documents the
IR system should return to that user. From a geometric perspective,
a pure IN is\emph{ }answered by a document with a probability that
depends on the length of the projection of the pure IN vector onto
the document subspace. Because of the uncertainty attached to the
IR search process, we suppose that the information being searched
by a user can be represented by a set of such pure INs, one for each
possible \emph{pure }IN that composes a user's IN.

To compute a probability of relevance of a document to a user's IN,
we make use of the generalisation of probabilities developed in quantum
physics, which is strongly connected to the geometry of the space
used to represent events and densities. A probabilistic event is represented
as a subspace (denoted $S$) in a Hilbert space%
\footnote{Hilbert spaces (roughly, vector spaces with complex scalars) are a
central mathematical concept in quantum physics.%
}. Let us assume that $S$ is the event {}``the document is relevant''.
A probability can first be defined for a pure IN, represented as a
\emph{unit} vector $\varphi$, by computing the length of the projection
of the vector $\varphi$ onto the subspace $S$, that is by computing
the value $\left\Vert \widehat{S}\varphi\right\Vert ^{2}$ where $\widehat{S}$
is the projector onto the subspace~$S$. This value is the probability
that the document is relevant with respect to the pure IN%
\footnote{We have $\left\Vert \widehat{S}\varphi\right\Vert ^{2}\in\left[0,1\right]$
since $\left\Vert \varphi\right\Vert =1$.%
}. 

When a user starts interacting with an IR system by, for instance,
typing a query%
\footnote{Queries are what (usually) users provide to an IR system, as means
to express their INs~\cite{Ingwersen2005The-Turn-Integration}.%
}, we first compute~(see Section~\ref{sub:document-representation})
an initial set of weighted pure IN vectors, where each weight is the
probability that the pure IN corresponds to the actual user's IN.
This captures the uncertainty typical to IR where firstly, the representation
is only an approximation of the user's IN, and, secondly, the query
may be ambiguous. The goal of an IR system is to reduce this undeterminism
through interaction. 

More formally, we assume that each pure IN vector $\varphi_{i}$ is
associated with a probability $p_{i}$~(the weight). We define the
probability of the event $S$ by using the usual total probability
theorem (across all possible pure INs)%
\footnote{As in quantum physics, we assume different $\varphi_{i}$ correspond
to different systems and are thus mutually exclusive.%
}:\begin{eqnarray}
\pr\left(S\right) & = & \sum_{i}p_{i}\pr\left(S|\varphi_{i}\right)=\sum_{i}p_{i}\braAket{\varphi_{i}}{\widehat{S}}{\varphi}_{i}=\tr\left(\rho\widehat{S}\right)\label{eq:trace-pr}\end{eqnarray}
where $\tr$ is the trace operator~\cite[p. 83]{Rijsbergen2004The-Geometry-of-Information}
and $\rho=\sum_{i}p_{i}\varphi_{i}\varphi_{i}^{\top}$ is called a
\emph{density operator}%
\footnote{We will omit the term {}``operator'' in the remaining of the paper.%
} and corresponds to a (probabilistic) \emph{mixtur}e of the pure INs
$\varphi_{i}$.\emph{ }In general, any operator $\rho$ characterised
by the fact that it is both positive-semi-definite%
\footnote{This means $v^{\top}\rho v\ge0$ for any vector $v$.%
} and of trace 1 defines a probability distribution over the subspaces,
i.e.~it is possible to interpret $\pr\left(S\right)=\mbox{tr}\left(\rho\widehat{S}\right)$
as a probability~\cite{Rijsbergen2004The-Geometry-of-Information}. 

For each document $d$, we compute a projector $\widehat{S}_{d}$
(Section~\ref{sub:document-representation}) and, for a query $q$,
the IN density $\rho$ is approximated by $\rho_{q}$~(Section~\ref{sub:Creating-the-Query}).
Using the projector $\widehat{S}_{d}$ and the density $\rho_{q}$,
the probability that a document $d$ is relevant to the query $q$
is then given by \foreignlanguage{english}{$\tr\left(\rho_{q}\widehat{S}_{d}\right)$}. 

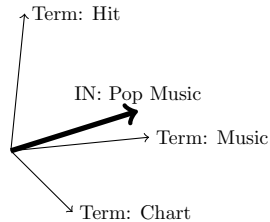
\begin{wrapfigure}{O}{0.5\columnwidth}%
\begin{centering}
% Sketch output, version 0.2 (build 161d, Tue Nov 24 16:28:42 2009)
% Output language: PGF/TikZ,LaTeX
\begin{tikzpicture}[line join=round]
\draw[draw=black,arrows=->](0,0)--(1.819,.181);
\draw[line width=2pt,arrows=->](0,0)--(1.672,.528);
\draw[draw=black,arrows=->](0,0)--(.181,1.819);
\draw[draw=black,arrows=->](0,0)--(.811,-.811);
\draw[anchor=south] (1.672,.528) node[scale=0.75] {IN: Pop Music};
\draw[anchor=west] (.181,1.819) node[scale=0.75] {Term: Hit};
\draw[anchor=west] (.811,-.811) node[scale=0.75] {Term: Chart};
\draw[anchor=west] (1.819,.181) node[scale=0.75] {Term: Music};
\end{tikzpicture}% End sketch output
\par\end{centering}

\caption{\label{fig:in-termspace}A pure IN in a term space}
\end{wrapfigure}

In our work, we assume that the vector space of pure INs is the term
space, where each dimension corresponds to a term. A pure IN is hence
described by a series of weighted terms. A (simplified) example is
shown in Figure~\ref{fig:in-termspace}, where the pure IN {}``pop
music''~(one unit vector) is represented by the terms {}``music'',
{}``chart'' and {}``hit'' of the term space. We show now how document
and query representations are computed in this term space.

\section{Creating the Document Subspace}

\label{sub:document-representation}

It is reasonable to assume that a typical document answers various
(pure) INs, since it is likely to contain answers~(be relevant) to
several queries. Moreover, \cite{Piwowarski2009Sound-and-Complete}~have
shown in the context of XML retrieval, that answers to topics~(statements
of INs) usually correspond to document fragments and not full documents.
Building on this, we assume that for each document there is a mapping
between its~(possibly overlapping and non-contiguous) fragments and
a set of pure INs. 

A document is thus associated with a set $\mathcal{U}_{d}$ of vectors
in the IN space. We hypothesise that a document is \emph{fully} relevant
to a pure IN if the latter can be written as a linear combination
of the vectors of $\mathcal{U}_{d}$, that is, if it is contained
in the subspace $S_{d}$ defined as the span of the vectors in $\mathcal{U}_{d}$.
The document will be \emph{partially} relevant to a pure IN with a
probability that depends on the length of the projection of the pure
IN vector onto the subspace $S_{d}$. The subspace $S_{d}$ can be
interpreted as a geometric representation of the event {}``the document
is relevant''. This construction process was validated in a document
filtering task~\cite{Piwowarski2010Filtering-documents}. In this paper, we investigate the effect of several parameters~(written
in bold below) on this process.

\textbf{Document Fragments. }We now assume that document fragments
are disjoint, and are obtained through a {}``natural'' segmentation
of the document. Various choices are possible, and our first strategy
is to use a single fragment, the document itself. This corresponds
to the vector space approach where a single vector represents a document.
The second strategy is to use paragraphs as fragments as they seem
to be of an appropriate size to correspond to a pure\emph{ }IN. We
also selected a third type of fragment, the sentence, as it is one
of the smallest coherent units in a document.

\textbf{Weighting Schemes. }We now need a vector representation for
each fragment. Three weighting schemes are used, namely, tf-idf, tf
and binary~(term presence/absence). The latter two are chosen since
they allow substantial reduction in computational complexity. In addition,
binary vectors are close or equal to tf vectors for small fragments,
for example, sentences.

$\mathcal{U}_{d}$ is formally defined as the set of vectors associated
with a document $d$, obtained through one of the above segmentation
and weighting scheme, i.e., we have one vector for each fragment.
As discussed before, we need to compute the subspace $S_{d}$ spanned
by the vectors of $\mathcal{U}_{d}$. For this, we use an eigenvalue
decomposition where \textbf{$\sum_{\varphi\in\mathcal{U}_{d}}\ket{\varphi}\bra{\varphi}$
}is expressed as $\sum_{i=1}^{D}\lambda_{i}v_{i}v_{i}^{\top}$ where
$D$ is the number of eigenvectors with non null eigenvalues~($D$
is also the dimension of the associated subspace), $\lambda_{i}>0$
are the eigenvalues~(we suppose without loss of generality that they
are of decreasing magnitude, i.e. $\lambda_{i}\ge\lambda_{i+1}$)
and the vectors $v_{i}$ form an orthonormal basis of the subspace
$S_{d}$~\cite{Stewart2001Eigensystems}. 

\textbf{Dimension selection. }As the vectors constructed from the
terms occurring in the document fragments are only an approximation
of the underlying pure IN vectors, the vectors from $\mathcal{U}_{d}$
will contain terms that should not be associated with the document.
We are thus interested in the eigenvectors associated with the $K$
highest eigenvalues since low eigenvalues are likely to be associated
with noise~\cite{Efron2005Eigenvalue-based-model}. We are interested
in measuring the effect of different dimensions to represent a document.
Hence, we chose a simple strategy, where we keep the eigenvectors
whose eigenvalue is higher than the average of the eigenvalues, which
we compared to two extreme strategies, namely, the case where we select
the eigenvector with the highest eigenvalue~(one dimension, $K=1$)
and the case where we keep all the eigenvectors~(full dimension,
$K=D$).

Finally, the projector $\widehat{S}_{d}$ associated with the $K$
dimensional subspace of document $d$ is expressed as $\sum_{i=1}^{K}v_{i}v_{i}^{\top}$.

\section{Creating the Query Density}

\label{sub:Creating-the-Query}

We now focus on the primary contribution of the paper, namely, the
construction of the IN density $\rho_{q}$ for a given query $q$.

As a query in its simplest form consists of a set of terms, we are
first interested in building the query representation for a query
composed of a single term, $t$. We described how a document is represented
as a set of pure IN vectors corresponding to different fragments of
the document. We extend this idea, and suppose that a query term $t$
can be represented as the set $\mathcal{U}_{t}$ of pure IN vectors
that correspond to document fragments containing the term $t$. That
is, we use the immediate surroundings of the term occurrences in the
documents of the collection being searched to build that term representation.
This is similar to pseudo-relevance feedback using passages from retrieved
documents containing the query terms~\cite{Allan1995Relevance-feedback}.
The difference is that we use all the passages to build the query
representation as we want to consider all possible pure INs associated
with the term $t$.

As we have \emph{a priori} no way to distinguish between the different
vectors in $\mathcal{U}_{t}$, we assume that each vector is equally
likely to be a pure IN composing the user's actual IN. Hence, a document
is relevant to the user's IN if it is relevant to any of the vectors
of $\mathcal{U}_{t}$, where the vectors are drawn with a uniform
probability. The corresponding density is then written as: \begin{equation}
\rho_{t}=\frac{1}{N_{t}}\sum_{\varphi\in\mathcal{U}_{t}}\varphi\varphi^{\top}\label{eq:mixture-term}\end{equation}
where $N_{t}$ is the number of vectors associated with term $t$
(the cardinality of $\mathcal{U}_{t}$). This definition of $\rho_{t}$
has all the required properties of a density~(see Section~\ref{sec:framework}).
In practice, this representation of a single-term query $t$ means
that, the more vectors $\varphi$ from $\mathcal{U}_{t}$ lie in the
document subspace, the higher the relevance of the document to the
query. This query representation hence favours documents containing
different {}``aspects'' of the IN, each of them as represented by
one of the pure INs in $\mathcal{U}_{t}$ associated with a query
term $t$. 

We discuss next the representation of a query composed of several
terms. There are three main parameters (written in bold below).

\textbf{Weighting scheme.} As for documents, three weighting schemes,
namely, tf-idf, tf and binary, are used to build the vectors forming
$\mathcal{U}_{t}$.

\textbf{Query construction~(mixture). }The above query representation~(Equation~\ref{eq:mixture-term})
can be generalised to a query composed of several terms. We assume
that a relevant document should equally answer all pure INs associated
with\emph{ }each query term. To compute the probability of relevance
of a document $d$, we first select a term from the query~(with a
probability $w_{t}$, see the next paragraph), and then one of the
vectors in $\mathcal{U}_{t}$. With this vector, we compute the probability
of document $d$ to be relevant to this pure IN. We repeat the process
and average over all the possible combinations. This defines the probability
of relevance of document $d$ given the query. Formally, this corresponds
to a density defined as a \emph{mixture} of all the pure IN vectors
associated with the query terms. This density is built from the individual
query term densities $\rho_{t}$ (Equation \ref{eq:mixture-term}):
\begin{equation}
\rho_{q}^{(m)}=\sum_{t\in q}\sum_{\varphi\in\mathcal{U}_{t}}\frac{w_{t}}{N_{t}}\varphi\varphi^{\top}=\sum_{t\in q}w_{t}\rho_{t}\label{eq:mixture}\end{equation}

\textbf{Query term weight. }The weights $w_{t}$ are used to quantify
the importance of each term $t$ of the query. We experimented with
two settings, one where all the $w_{t}$ were equal, and the other
where they were set to the corresponding term idf values. In both
approaches, we normalise the weights so their sum equals 1.

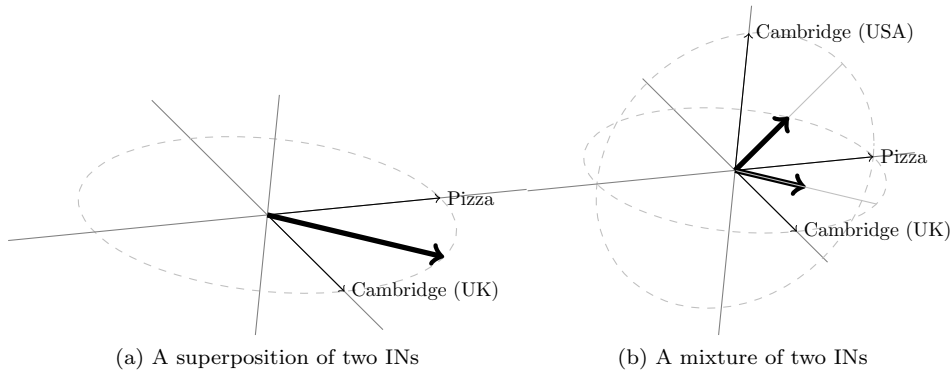
\begin{figure*}
\begin{centering}
\subfloat[A superposition of two INs]{% Sketch output, version 0.2 (build 160d, Fri Aug 28 09:11:05 2009)
% Output language: PGF/TikZ,LaTeX
\begin{tikzpicture}[line join=round]
\draw[draw=gray](-1.521,1.521)--(1.521,-1.521);
\draw[color=lightgray,dashed](2.274,.226)--(2.162,.326)--(2.028,.422)--(1.874,.515)--(1.702,.602)--(1.512,.683)--(1.308,.758)--(1.09,.825)--(.862,.884)--(.625,.934)--(.382,.974)--(.135,1.005)--(-.113,1.026)--(-.36,1.037)--(-.604,1.038)--(-.841,1.028)--(-1.07,1.008)--(-1.289,.978)--(-1.495,.938)--(-1.685,.889)--(-1.859,.831)--(-2.015,.765)--(-2.151,.691)--(-2.265,.611)--(-2.357,.524)--(-2.425,.432)--(-2.469,.336)--(-2.489,.236)--(-2.484,.134)--(-2.454,.031)--(-2.4,-.072)--(-2.322,-.175)--(-2.221,-.276)--(-2.097,-.375)--(-1.953,-.469)--(-1.79,-.559)--(-1.609,-.644)--(-1.412,-.722)--(-1.2,-.792)--(-.977,-.855)--(-.744,-.91)--(-.504,-.955)--(-.259,-.991)--(-.011,-1.017)--(.237,-1.033)--(.483,-1.039)--(.723,-1.034)--(.957,-1.019)--(1.181,-.994)--(1.393,-.959)--(1.592,-.915)--(1.775,-.861)--(1.94,-.799)--(2.085,-.729)--(2.211,-.652)--(2.314,-.568)--(2.394,-.479)--(2.45,-.384)--(2.482,-.286)--(2.489,-.186)--(2.472,-.083)--(2.43,.021)--(2.364,.124)--(2.274,.226)--(2.162,.326);
\draw[draw=gray](-3.411,-.339)--(3.411,.339);
\draw[draw=black,arrows=->](0,0)--(2.274,.226);
\draw[draw=gray](-.158,-1.592)--(.158,1.592);
\draw[draw=black,arrows=->](0,0)--(1.014,-1.014);
\draw[line width=2pt,arrows=->](0,0)--(2.325,-.557);
\draw[anchor=west] (1.014,-1.014) node[scale=0.75] {Cambridge (UK)};
\draw[anchor=west] (2.274,.226) node[scale=0.75] {Pizza};
\end{tikzpicture}% End sketch output
}\subfloat[A mixture of two INs]{% Sketch output, version 0.2 (build 160d, Fri Aug 28 09:11:05 2009)
% Output language: PGF/TikZ,LaTeX
\begin{tikzpicture}[line join=round]
\draw[draw=gray](-1.217,1.217)--(1.217,-1.217);
\draw[color=lightgray,dashed](1.819,.181)--(1.729,.261)--(1.622,.338)--(1.499,.412)--(1.361,.482)--(1.21,.547)--(1.046,.606)--(.872,.66)--(.69,.707)--(.5,.747)--(.306,.78)--(.108,.804)--(-.09,.821)--(-.288,.83)--(-.483,.83)--(-.673,.822)--(-.856,.806)--(-1.031,.782)--(-1.196,.751)--(-1.348,.711)--(-1.488,.665)--(-1.612,.612)--(-1.721,.553)--(-1.812,.489)--(-1.885,.419)--(-1.94,.346)--(-1.975,.269)--(-1.991,.189)--(-1.987,.107)--(-1.963,.025)--(-1.92,-.058)--(-1.857,-.14)--(-1.776,-.221)--(-1.678,-.3)--(-1.563,-.375)--(-1.432,-.447)--(-1.287,-.515)--(-1.129,-.577)--(-.96,-.634)--(-.782,-.684)--(-.596,-.728)--(-.403,-.764)--(-.207,-.793)--(-.009,-.814)--(.19,-.827)--(.386,-.831)--(.579,-.827)--(.766,-.815)--(.945,-.795)--(1.115,-.767)--(1.274,-.732)--(1.42,-.689)--(1.552,-.639)--(1.668,-.583)--(1.768,-.521)--(1.851,-.454)--(1.915,-.383)--(1.96,-.308)--(1.986,-.229)--(1.991,-.148)--(1.977,-.066)--(1.944,.017)--(1.891,.099)--(1.819,.181)--(1.729,.261);
\draw[color=lightgray,dashed](1.819,.181)--(1.828,.361)--(1.819,.538)--(1.792,.709)--(1.747,.873)--(1.684,1.029)--(1.605,1.174)--(1.51,1.308)--(1.4,1.429)--(1.276,1.535)--(1.139,1.626)--(.991,1.701)--(.833,1.759)--(.667,1.8)--(.494,1.823)--(.316,1.828)--(.135,1.814)--(-.047,1.783)--(-.228,1.733)--(-.408,1.667)--(-.583,1.584)--(-.753,1.485)--(-.915,1.371)--(-1.068,1.244)--(-1.211,1.105)--(-1.341,.954)--(-1.458,.794)--(-1.561,.626)--(-1.648,.452)--(-1.718,.274)--(-1.772,.092)--(-1.808,-.09)--(-1.826,-.271)--(-1.826,-.45)--(-1.808,-.624)--(-1.771,-.792)--(-1.717,-.952)--(-1.647,-1.103)--(-1.559,-1.243)--(-1.457,-1.37)--(-1.339,-1.484)--(-1.209,-1.583)--(-1.066,-1.666)--(-.913,-1.733)--(-.751,-1.782)--(-.581,-1.814)--(-.406,-1.828)--(-.226,-1.823)--(-.044,-1.801)--(.138,-1.76)--(.319,-1.702)--(.496,-1.627)--(.669,-1.536)--(.835,-1.43)--(.993,-1.31)--(1.141,-1.176)--(1.277,-1.031)--(1.401,-.875)--(1.511,-.711)--(1.606,-.54)--(1.685,-.363)--(1.747,-.183)--(1.792,-.001)--(1.819,.181)--(1.828,.361);
\draw[draw=gray](-2.729,-.271)--(2.365,.235);
\draw[draw=gray](-.217,-2.183)--(.217,2.183);
\draw[draw=black,arrows=->](0,0)--(1.819,.181);
\draw[draw=lightgray](0,0)--(1.414,1.414);
\draw[line width=2pt,arrows=->](0,0)--(.707,.707);
\draw[line width=2pt,arrows=->](0,0)--(.93,-.223);
\draw[draw=lightgray](0,0)--(1.86,-.446);
\draw[draw=black,arrows=->](0,0)--(.181,1.819);
\draw[draw=black,arrows=->](0,0)--(.811,-.811);
\draw[anchor=west] (.181,1.819) node[scale=0.75] {Cambridge (USA)};
\draw[anchor=west] (.811,-.811) node[scale=0.75] {Cambridge (UK)};
\draw[anchor=west] (1.819,.181) node[scale=0.75] {Pizza};
\end{tikzpicture}% End sketch output
}
\par\end{centering}

\caption{\label{fig:examples}Combining INs}

\end{figure*}

We present a second query construction process, inspired from IR and
quantum theory. In vectorial IR, a query is represented by a vector
that corresponds to a linear combination of the vectors associated
with the query terms. In quantum theory, a normalised linear combination
corresponds to the principle of superposition, where the description
of a system state can be \emph{superposed} to describe a new system
state.

In our case, the system state corresponds to the user's pure IN, and
we use the superposition principle to build new pure INs from existing
ones, as illustrated with the example shown in Figure~\ref{fig:examples}.
Let $\ket{\varphi_{p}}$, $\ket{\varphi_{c/uk}}$ and $\ket{\varphi_{c/usa}}$
be three vectors in a three-dimensional IN space that, respectively,
represent the INs {}``I want a pizza'', {}``I want it to be delivered
in Cambridge~(UK)'' and {}``I want it to be delivered in Cambridge~(USA)''.
The pure IN vector {}``Pizza delivered in Cambridge~(UK)'' would
be represented by a (normalised) linear combination~(or superposition)
of $\varphi_{p}$ and $\varphi_{c/uk}$, as depicted in Figure \ref{fig:examples}(a).
We can similarly build the IN for Cambridge (USA). To represent the
ambiguous query {}``pizza delivery in Cambridge'' where we do not
know whether Cambridge is in the USA or the UK, and assuming there
is no other source of ambiguity, we would use a mixture of the two
possible superposed INs, as depicted by the two vectors of the mixture
in Figure~\ref{fig:examples}(b), which brings us to another variant
of query construction, the mixture of superpositions.

\textbf{Query construction~(mixture of superpositions). }To compute
the probability of relevance, for each term $t$ of the query, we
randomly select a vector from the set $\mathcal{U}_{t}$. We then
superpose~(i.e., compute a linear combination) the selected vectors~(one
for each term), where the weight in the linear combination is $\sqrt{w_{t}}$~(see
below for why we use a square root). From this vector, we compute
the probability of the document to be relevant to this IN made from
the superposition of IN vectors~(one per query term). With respect
to our example, the set $\mathcal{U}_{pizza}$ would be just one vector~({}``\emph{I
want a pizza to be delivered''}), and $\mathcal{U}_{Cambridge}$
would contain two vectors~(one for UK, one for USA). 

As with the simple mixture approach, the above process can be repeated
for all the possible selections of vectors and the corresponding query
density is:

\begin{equation}
\rho_{q}^{(ms)}=\frac{1}{Z_{q}}\sum_{\varphi_{1}\in\mathcal{U}_{t_{1}}}\cdots\sum_{\varphi_{n}\in\mathcal{U}_{t_{n}}}\left(\sum_{i=1}^{n}\sqrt{\frac{w_{t_{i}}}{N_{t_{i}}}}\ket{\varphi_{i}}\right)\left(\sum_{i=1}^{n}\sqrt{\frac{w_{t_{i}}}{N_{t_{i}}}}\ket{\varphi_{i}}\right)^{\top}\label{eq:mixture-superposition}\end{equation}
where $Z_{q}$ is a normalisation coefficient, and $t_{i}\,(i=1\ldots n)$
are the $n$ query terms. We use $N_{t}$ to ensure that each term
contribution is equally important, and square roots because both $N_{t}$
and $w_{t}$ appear two times in the above formula. In theory the
vector $\sum_{i}\sqrt{\frac{w_{t_{i}}}{N_{t_{i}}}}\ket{\varphi_{i}}$
should be normalised but to obtain a computable formula we did not
do so %
\footnote{The effect will be to give higher importance to superpositions of
vectors $\varphi_{i}$ who are similar, i.e., whose cosine is closer
to 1. %
}.

Note that for one-term queries, the two described query constructions~(mixture
and mixture of superpositions) give the same result. Another important
point from a computational perspective is that in both cases, the
query can be estimated from single term densities~(not demonstrated
for Equation \ref{eq:mixture-superposition}). We hence pre-compute
the densities $\rho_{t}$ for each term $t$, and use them at query
time to compute $\rho_{q}^{(m)}$ and $\rho_{q}^{(ms)}$. 

\textbf{Dimension selection.} As for the representation of documents,
both densities are expressed, through eigenvalue decomposition, as
a sum $\sum_{i=1}^{D}\lambda_{i}v_{i}v_{i}^{\text{\ensuremath{\top}}}$
where the $(\lambda_{i},v_{i})$ are eigenpairs ordered by decreasing
eigenvalues. Our final density used for computing the probability
of relevance is then $\rho_{q}=\sum_{i=1}^{K}\lambda_{i}v_{i}v_{i}^{\top}$
where $K$ is the selected dimension~(where $K\le D$). We use the
same three strategies to set $K$ that were used for the document
representations (see end of Section~\ref{sub:document-representation}).

\section{Experiments and Analysis}

\label{sec:experiments}

In previous work~\cite{Piwowarski2010Filtering-documents}, we validated
the subspace document representation on a filtering task. In this
paper, we explore both the document and the query representations
in an ad hoc retrieval task. In particular, we look at the effects
of the parameters discussed in Sections~\ref{sub:document-representation}
and~\ref{sub:Creating-the-Query}. These are listed on the left column
of Table~\ref{tab:Main-result}. As the parameters are mostly independent
from each other, we experimented with 756 settings; those not making
sense were ignored%
\footnote{When using a whole document as fragment, the document subspace is
one-dimensional and in this case there is no point to investigate
the dimension selection parameter.%
}.

We used the INEX 2008 collection in our experiments because its documents
have markup (in XML format) delineating text units. The collection
consists of 659,388 Wikipedia documents in XML format, using tags
such as article, section and paragraph to model a document logical
structure~\cite{Denoyer2006The-Wikipedia-XML-Corpus}. INEX 2008
has 70 assessed topics, and for each topic, relevant passages in (pooled)
documents were highlighted by human assessors. A document containing
a relevant passage is assumed relevant, which is in accordance with
\textsc{Trec} guidelines.

We preprocessed the documents by extracting the fragments, i.e., the
whole document, the paragraphs~(as determined by the XML markup)
and the sentences%
\footnote{We use \url{http://www.andy-roberts.net/software/jTokeniser/index.html}
for this.%
}. We then stemmed and stopped~(using the SMART list of stop-words)
the text fragments. For each term $t$, we computed an approximation
of the term density $\rho_{t}$ (Equation~\ref{eq:mixture-term})
based on a sample of 10,000 documents~(maximum) containing the term
$t$ and using a thin eigenvalue decomposition with maximum rank set
to 10~\cite[pp. 171-181]{Stewart2001Eigensystems}. This value, chosen
through experimentation, represents a good trade-off between complexity
and efficiency. For each query $q$, we computed the query density
$\rho_{q}$ using the densities $\rho_{t}$ of its composing terms
$t$, using either the simple mixture (Equation~\ref{eq:mixture})
or the mixture of superpositions (Equation~\ref{eq:mixture-superposition}).
Then, we first retrieved a set of 1,500 documents using BM25%
\footnote{With the standard parameter values.%
}~\cite{Walker1999Okapi/Keenbow-at-TREC-8}. For each retrieved document
$d$ and each parameter setting, we computed the projector $\widehat{S}_{d}$
and computed a probability of relevance as $\mbox{tr}\left(\rho_{q}\widehat{S}_{d}\right)$.
We used this value to re-rank the documents.

\begin{table}
\begin{centering}
\begin{tabular}{|>{\centering}p{0.3\columnwidth}|>{\centering}p{0.6\columnwidth}|}
\hline 
\textbf{Parameters} & \textbf{Means}\tabularnewline
\hline
\hline 
{\footnotesize (1) Document fragment } & {\footnotesize sentence (0.14) >\textcompwordmark{}> paragraph (0.12)
>\textcompwordmark{}> document (0.11)}\tabularnewline
\hline 
{\footnotesize (2) Weighting scheme (document fragment)} & {\footnotesize tf (0.13) >\textcompwordmark{}> tf-idf (0.12), binary
(0.12)}\tabularnewline
\hline 
{\footnotesize (3) Weighting scheme (query)} & {\footnotesize tf-idf (0.13) >\textcompwordmark{}> tf (0.12), tf-idf
> binary (0.12)}\tabularnewline
\hline 
{\footnotesize (4) Dimension selection (document)} & {\footnotesize all (0.14) >\textcompwordmark{}> highest (0.11), mean
(0.14) >\textcompwordmark{}> highest (0.11)}\tabularnewline
\hline 
{\footnotesize (5) Dimension selection (query)} & {\footnotesize all (0.13), mean (0.13), highest (0.12)}\tabularnewline
\hline 
{\footnotesize (6) Term weight in query} & {\footnotesize idf (0.13) >\textcompwordmark{}> uniform (0.12)}\tabularnewline
\hline 
{\footnotesize (7) Query construction} & {\footnotesize mixture (0.13), mixture of superpositions (0.13)}\tabularnewline
\hline
\end{tabular}
\par\end{centering}

\caption{\label{tab:Main-result}Means of medians of average precision for
each topic. The {}``>'' (resp. {}``>\textcompwordmark{}>'') sign
is used to denote statistical significance at 0.05 (resp. 0.01).}

\end{table}

Table~\ref{tab:Main-result} shows our results. For each parameter
(left column), we show in the right column the means of the medians
of average precision computed for the different settings of that parameter.
For example in row (1), when the fragment is {}``sentence'', this
value is 0.14. To compare two settings, say {}``sentence'' vs.~{}``paragraph'',
we performed a paired t-test where each pair of samples corresponds
to the same topic and same parameter values~(weighting scheme, dimension
selection, query term weight, query construction) but for the document
fragment setting. For this example, the result shows that using sentence
fragments outperformed paragraph fragments at a 0.01 significance
level. We discuss each result next.

For the document fragment parameter (1), the best performing setting
was with {}``sentence'' followed by {}``paragraph'' and {}``document''.
Each time the difference was found to be significant at a 0.01 level.
This indicates that the right level of segmentation (to construct
the pure IN vectors) is at sentence level.

Overall, the weighting scheme for document fragments and queries had
some effect on retrieval effectiveness. For building the query term
density~(3), the tf-idf scheme led to significantly better results,
whereas for document fragments~(2), the tf scheme performed better.
The results are somehow in contradiction with vectorial IR findings,
but might stem from the fact that to build the query term representation
we sample much more vectors than for the document one; hence in the
former case it is important to weight terms according to their importance
(idf). When looking more in details into the results, we also found
out that the weighting scheme was highly dependent on the other parameters,
and should hence be chosen depending on them. 

The setting of the subspace dimension has a different effect on documents
and queries. For documents~(4), performance was improved using the
full dimension or the mean of the eigenvalues (to determine the dimension
of the subspace representation). This shows that using more than one
dimension to represent a document is beneficial. However, for queries~(5)
we observe only a slight improvement when using multiple dimensions~(none
of which were significant). 

\begin{table}
\begin{tabular}{|c||>{\centering}p{0.35\columnwidth}||c|>{\centering}p{0.38\columnwidth}|}
\hline 
\multicolumn{2}{|c||}{\textbf{\footnotesize Mixture of superpositions}} & \multicolumn{2}{c|}{\textbf{\footnotesize Mixture}}\tabularnewline
\hline 
{\footnotesize $\Delta_{AP}$} & \textbf{\footnotesize Topic} & {\footnotesize $\Delta_{AP}$} & \textbf{\footnotesize Topic}\tabularnewline
\hline 
{\footnotesize 0.22} & {\footnotesize social networks mining} & {\footnotesize 0.32} & {\footnotesize \textquotedbl{}records management\textquotedbl{} metadata}\tabularnewline
\hline 
{\footnotesize 0.19} & {\footnotesize virtual museums} & {\footnotesize 0.16} & {\footnotesize Tata Motors Company in India}\tabularnewline
\hline 
{\footnotesize 0.10} & {\footnotesize genetically modified food safety} & {\footnotesize 0.15} & {\footnotesize Nikola Tesla inventions patents}\tabularnewline
\hline 
{\footnotesize 0.09} & {\footnotesize wikipedia vandalism} & {\footnotesize 0.08} & {\footnotesize vodka producing countries}\tabularnewline
\hline 
{\footnotesize 0.06} & {\footnotesize flower meaning} & {\footnotesize 0.08} & {\footnotesize mahler symphony song}\tabularnewline
\hline
\end{tabular}

\caption{Top five performing topics using, respectively, mixture of superpositions
(Equation \ref{eq:mixture-superposition}) and mixture (Equation \ref{eq:mixture})
as query representation.\label{tab:ms-vs-m}}

\end{table}

For the query construction methodology, we first see that weighting
the query terms by their idf values outperformed using a uniform scheme~(6).
When looking at a mixture vs.~mixture of superpositions~(7), no
significant overall performance difference exists. However, we observe
different behaviours depending on the topic. Table~\ref{tab:ms-vs-m}
shows the best performing topics for, respectively, the mixture of
superpositions and the mixture. The topics better handled by the mixture
of superpositions are topics for which the terms form a {}``concept'',
for example {}``social networks mining'' where the three terms together
have a specific meaning. For the mixture, topics for which each term
reflects a different aspect of the topic, e.g. {}``\textquotedbl{}records
management\textquotedbl{} metadata'', where {}``metadata'' and
{}``records management'' are the two different concepts, had a better
performance. This indicates that selecting the query density computation
according to the topic may prove beneficial.

The above example suggests that it may be beneficial to treat parts
of the query differently by combining both construction methods into
one query. For example, the terms {}``records'' and {}``management''
form a single aspect and should thus be superposed. Afterwards, the
superposed terms should be mixed with {}``metadata'', which describes
another aspect, to answer the query {}``\textquotedbl{}records management\textquotedbl{}
metadata''. In general, to determine which terms form a single concept,
we can rely on explicit markers like quotes in this example, or on
an automatic algorithm based e.g. on co-occurrences.

\begin{wrapfigure}{O}{0.45\columnwidth}%
\begin{centering}
\includegraphics[scale=0.29]{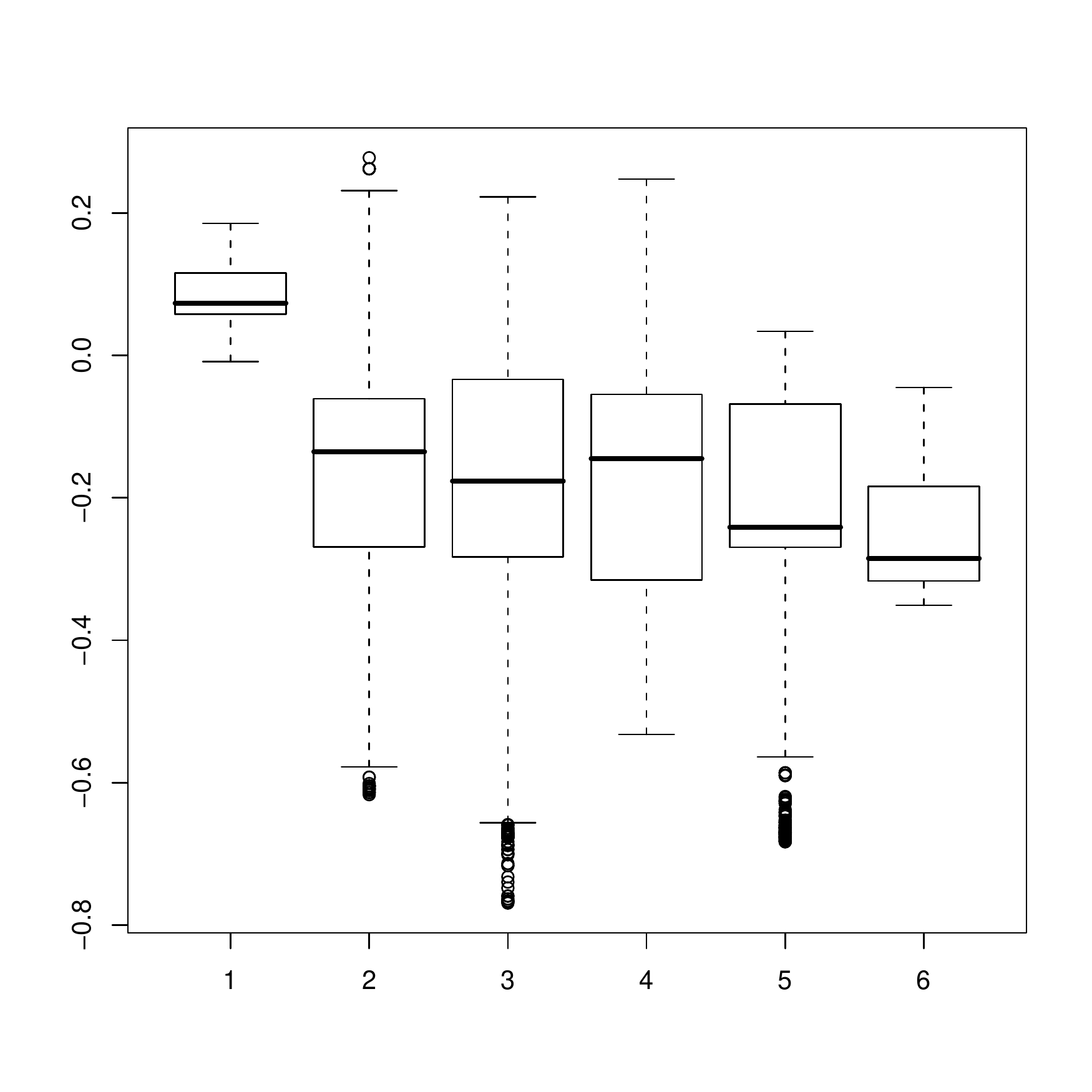}
\par\end{centering}

\caption{\label{fig:query-length}Boxplot of the effect of query length (number
of terms) on average precision. The x-axis is the query length~(number
of terms) and the y-axis is the difference in average precision between
BM25 and our method in different settings.}
\end{wrapfigure}%

We also compared our results to a state-of-the-art retrieval IR system,
namely BM25~\cite{Walker1999Okapi/Keenbow-at-TREC-8}. We found that
the performances of our framework were consistently lower in average~(using
standard IR evaluation metrics). A brief analysis~(not reported here)
comparing the results of the best performing configurations with BM25
for the topics in Table~\ref{tab:ms-vs-m} reveals that we could
get closer to BM25 performances by (again) choosing the right query
construction methodology~(mixture vs mixture of superpositions).

Finally, we investigated the effect of query length~(number of terms)
and the number of relevant documents (of a query) on retrieval effectiveness.
No correlation was found between the difference in performance between
BM25 and our framework, and the number of relevant documents. There
was however a strong dependency on the query length. As illustrated
in Figure~\ref{fig:query-length}, when the query length is one~(there
is no difference between the two query density construction methods),
our approach outperforms consistently BM25; when the number of terms
in the query increases, retrieval performance drops. This further
confirms that the appropriate calculation of the query density --
in particular for multi-term queries -- needs to be investigated.

\section{Conclusion and Future Work}

In this paper, we presented a methodology to build multidimensional
representations of documents and queries. These representations are
inspired from the geometric/probabilistic framework of quantum physics.
The latter allows us to compute probabilities of relevance based on
a more complex representations of documents than a simple bag of words,
namely, a multidimensional one based on document fragments. We believe
that such a multidimensional representation is key to a successful
framework for exploiting user's interaction~\cite{Rijsbergen2004The-Geometry-of-Information}.

We performed experiments to explore various parameters influencing
the effectiveness of our representations. We showed that using more
than one dimension to represent documents improves performance, confirming
previous results. Considering a document as a fragment, as done in
most classical models, is not sufficient to distinguish between the
different information needs a document covers. Indeed, while most
of the classical models only take the mere occurrence of a term into
account, we showed in our experiments that the vicinity of terms~(the
fact that they appear in the same fragment) plays an important role. 

We also explored two different and principled ways to construct the
query representation. We have shown that queries whose terms define
a concept and those whose terms are more independent are better handled
by two different methods, respectively, the mixture of superpositions
and the (simple) mixture. This suggests that we can gain further improvements
if both strategies are applied together in an adaptive manner. This
is part of our future work. 

As our representation of queries and documents aims at tackling interactive
IR, this works validates our framework for the most common first interaction
step between a user and an IR system~--~a user typing a query. Exploiting
further interaction steps~(for example viewing or saving a document),
is also part of our future work.

\paragraph*{{\small Acknowledgements}}

{\small This research was supported by an Engineering and Physical
Sciences Research Council grant (Grant Number EP/F015984/2). Mounia
Lalmas is currently funded by Microsoft Research/Royal Academy of
Engineering.}{\small \par}

\bibliographystyle{abbrv}
\bibliography{IQIR-RIAO-arxiv}

\begin{thebibliography}{10}

\bibitem{Allan1995Relevance-feedback}
J.~Allan.
\newblock {Relevance feedback with too much data}.
\newblock In E.~A. Fox, P.~Ingwersen, and R.~Fidel, editors, {\em 18th ACM
  SIGIR conference}, Seattle, Washington, United States, 1995. {ACM}.

\bibitem{Baeza-Yates1999Modern-Information}
R.~Baeza-Yates and B.~Ribeiro-Neto.
\newblock {\em Modern Information Retrieval}.
\newblock Addison Wesley, New York, USA, 1999.

\bibitem{Che2006A-stereo-document}
L.~Che, J.~Zen, and N.~Tokud.
\newblock {A "stereo" document representation for textual information
  retrieval}.
\newblock {\em JASIST}, 5, 2006.

\bibitem{Denoyer2006The-Wikipedia-XML-Corpus}
L.~Denoyer and P.~Gallinari.
\newblock {T}he {W}ikipedia {X}{M}{L} {C}orpus.
\newblock {\em SIGIR Forum}, 2006.

\bibitem{Efron2005Eigenvalue-based-model}
M.~Efron.
\newblock Eigenvalue-based model selection during latent semantic indexing.
\newblock {\em JASIST}, 56(9), 2005.

\bibitem{Ingwersen2005The-Turn-Integration}
P.~Ingwersen and K.~J{\"a}rvelin.
\newblock {\em The Turn: Integration of Information Seeking and Retrieval in
  Context (The Information Retrieval Series)}.
\newblock Springer-Verlag, Secaucus, NJ, USA, 2005.

\bibitem{Melucci2008A-basis-for-information}
M.~Melucci.
\newblock A basis for information retrieval in context.
\newblock {\em {ACM TOIS}}, 26(3), 2008.

\bibitem{Piwowarski2010Exploring-a-Multidimensional}
B.~Piwowarski, I.~Frommholz, M.~Lalmas, and K.~van Rijsbergen.
\newblock Exploring a multidimensional representation of documents and queries.
\newblock In {\em RIAO proceedings}, 2010.

\bibitem{Piwowarski2010Filtering-documents}
B.~Piwowarski, I.~Frommholz, Y.~Moshfeghi, M.~Lalmas, and K.~van Rijsbergen.
\newblock Filtering documents with subspaces.
\newblock In {\em Proceedings of the 32nd ECIR Conference}, 2010.
\newblock Poster.

\bibitem{Piwowarski2009A-Quantum-based-Model}
B.~Piwowarski and M.~Lalmas.
\newblock {A Quantum-based Model for Interactive Information Retrieval
  (extended version)}.
\newblock {\em ArXiv e-prints}, (0906.4026), 2009.

\bibitem{Piwowarski2009Sound-and-Complete}
B.~Piwowarski, A.~Trotman, and M.~Lalmas.
\newblock Sound and complete relevance assessments for {XML} retrieval.
\newblock {\em ACM TOIS}, 27(1), 2009.

\bibitem{Stewart2001Eigensystems}
G.~W. Stewart.
\newblock {\em Eigensystems}, volume~2 of {\em Matrix algorithms}.
\newblock SIAM, 2001.

\bibitem{Rijsbergen2004The-Geometry-of-Information}
C.~J. van Rijsbergen.
\newblock {\em The Geometry of Information Retrieval}.
\newblock Cambridge University Press, New York, NY, USA, 2004.

\bibitem{Walker1999Okapi/Keenbow-at-TREC-8}
S.~Walker and S.~E. Robertson.
\newblock Okapi/keenbow at {TREC}-8.
\newblock In E.~M. Voorhees and D.~K. Harman, editors, {\em NIST Special
  Publication 500-246: The Eighth Text REtrieval Conference (TREC-8)},
  Gaithersburg, Maryland, USA, 1999.

\bibitem{Wang2008A-study-of-methods}
X.~Wang, H.~Fang, and C.~Zhai.
\newblock A study of methods for negative relevance feedback.
\newblock In S.-H. Myaeng, D.~W. Oard, F.~Sebastiani, T.-S. Chua, and M.-K.
  Leong, editors, {\em Proceedings of the 31st Annual International ACM SIGIR},
  New York, NY, USA, 2008. ACM.

\bibitem{Widdows2003Orthogonal-negation}
D.~Widdows.
\newblock Orthogonal negation in vector spaces for modelling word-meanings and
  document retrieval.
\newblock In {\em Proceedings of the 41st ACL conference}, Morristown, NJ, USA,
  2003. Association for Computational Linguistics.

\end{thebibliography}

\end{document}